\documentclass[onecolumn]{aastex62}
\pdfoutput=1
\usepackage{graphicx}
\usepackage{amssymb}                    
\usepackage{ulem}
\usepackage{float}
\usepackage{rotating}
\usepackage{enumitem}
\usepackage{rotating}
\usepackage{makecell}
\usepackage{appendix}
\usepackage{fixmath}
\newcommand{\hic}{Hi-C 2.1}
\newcommand{\ar}{AR\,12712}

\newcommand{\arc}{\ensuremath{^{\prime\prime}}}

\usepackage{times} 
\usepackage{amsmath}
\usepackage{booktabs}

\received{24 September, 2019}
\accepted{20 January, 2020}
\submitjournal{ApJ}
\begin{document}
\shortauthors{Williams, Walsh, Winebarger (2019)}
\shorttitle{\hic\ Coronal Strands}

\title{Is the High-Resolution Coronal Imager Resolving Coronal Strands? Results from \ar}
\author{Thomas Williams}
\affil{Jeremiah Horrocks Institute, UCLan, Preston, PR1 2HE, UK}
\author{Robert W.\ Walsh}
\affil{Jeremiah Horrocks Institute, UCLan, Preston, PR1 2HE, UK}
\author{Amy R.\ Winebarger}
\affil{NASA Marshall Space Flight Center, ST13, Huntsville, AL 35812, USA}
\author{David H.\ Brooks}
\affil{College of Science, George Mason University, 400 University Drive, Fairfax, VA 22030, USA}
\author{Jonathan W.\ Cirtain}
\affil{BWX Technologies, Inc., 800 Main St \#400, Lynchburg, VA 24504, USA}
\author{Bart De Pontieu}
\affil{Lockheed Martin Solar and Astrophysics Laboratory, Palo Alto, CA 94304, USA}
\author{Leon Golub}
\affil{Harvard-Smithsonian Center for Astrophysics, 60 Garden St., Cambridge, MA 02138, USA}
\author{Ken Kobayashi}
\affil{NASA Marshall Space Flight Center, ST13, Huntsville, AL 35812, USA}
\author{David E.\ McKenzie}
\affil{NASA Marshall Space Flight Center, ST13, Huntsville, AL 35812, USA}
\author{Richard J.\ Morton}
\affil{Department of Mathematics, Physics and Electrical Engineering, Northumbria University, Newcastle Upon Tyne, NE1 8ST, UK}
\author{Hardi Peter}
\affil{Max Planck Institute for Solar System Research, Justus-von-Liebig-Weg 3, D-37077, G\"{o}ttingen, Germany}
\author{Laurel A.\ Rachmeler}
\affil{NASA Marshall Space Flight Center, ST13, Huntsville, AL 35812, USA}
\author{Sabrina L.\ Savage}
\affil{NASA Marshall Space Flight Center, ST13, Huntsville, AL 35812, USA}
\author{Paola Testa}
\affil{Smithsonian Astrophysical Observatory, 60 Garden Street, MS 58, Cambridge, MA 02138, USA}
\author{Sanjiv K.\ Tiwari}
\affil{Bay Area Environmental Research Institute, NASA Research Park, Moffett Field, CA 94035, USA}
\affil{Lockheed Martin Solar and Astrophysics Laboratory, Palo Alto, CA 94304, USA}
\author{Harry P.\ Warren}
\affil{Space Science Division, Naval Research Laboratory, Washington, DC 20375, USA}
\author{Benjamin J.\ Watkinson}
\affil{Jeremiah Horrocks Institute, UCLan, Preston, PR1 2HE, UK}

\begin{abstract}
Following the success of the first mission, the \textit{High-Resolution Coronal Imager} (Hi-C) was launched for a third time (\hic) on 29\textsuperscript{th} May 2018 from the White Sands Missile Range, NM, USA.  On this occasion, 329 seconds of 17.2 nm data of target active region \ar\ was captured with a cadence of $\approx4$ s, and a plate scale of 0.129 \arc\/pixel. Using data captured by \hic\ and co-aligned observations from SDO/AIA 17.1 nm we investigate the widths of 49 coronal strands. We search for evidence of substructure within the strands that is not detected by AIA, and further consider whether these strands are fully resolved by \hic. With the aid of Multi-Scale Gaussian Normalization (MGN), strands from a region of low-emission that can only be visualized against the contrast of the darker, underlying moss are studied. A comparison is made between these low-emission strands with those from regions of higher emission within the target active region. It is found that \hic\ can resolve individual strands as small as $\approx202$ km, though more typical strands widths seen are $\approx513$ km. For coronal strands within the region of low-emission, the most likely width is significantly narrower than the high-emission strands at $\approx388$ km. This places the low-emission coronal strands beneath the resolving capabilities of SDO/AIA, highlighting the need of a permanent solar observatory with the resolving power of Hi-C.
\end{abstract}

\section{Introduction}
On the 11\textsuperscript{th} July 2012 the NASA sounding rocket, Hi-C was first-launched and captured high resolution ($\approx 0.3-0.4$~\arc), high-cadence ($\approx 5$~s) images of active region 11520 \citep{kobayashi14} in a  narrowband 19.3 nm channel. The unprecedented capabilities of this instrument allowed the corona to be viewed in greater detail than previously capable by space-borne instruments, \textit{e.g.} SDO/AIA (1.5~\arc, 12~s; \citealp{lemen12}) and SoHO/EIT (5~\arc, 12~min; \citealp{delaboudiniere95}). Studies on the data obtained during the maiden flight of Hi-C revealed new information of the small-scale structures in the corona and transition region. Several publications have resulted from those five minutes of observations, including energy release along braided structures \citep{cirtain13,thalmann14,tiwari14,pontin17}, possible nanoflare heating in active region moss \citep{testa13,winebarger13}, coronal loop structure \citep{peter13,barczynski17,aschwanden17}, and counter-streaming along filament-channels \citep{alexander13}.

Coronal loops form one of the basic building blocks of the corona as they exist in both the quiet Sun and in active regions. Observational investigation of their structure has existed since the 1940s \citep{bray91} with the loops viewed in EUV and X-ray channels. For active regions, there are two loop types that have predominantly been studied: the short, hot loops in an active region core, typically observed in X-rays, and the cooler, longer loops that surround the core, typically observed in EUV \citep{reale10}. EUV loops are observed to evolve and cool. They are relatively steady over periods of several hours \citep{antiochos03,warren10,warren11}. In comparison, active region core loops are hotter, shorter and found in strong magnetic field areas within an active region \citep{berger99}. Typically, coronal loops have lengths of the same order as the barometric scale height, though there are suggestions of miniature loops in the chromosphere which span just a single granule \citep{feldman83,peter13,barczynski17}.

When studying the heating of coronal loops, one of the most important factors for consideration is whether the observed loop structure is isothermal or multi-thermal along the line of sight. If an observed loop is isothermal, it could indicate that the loop structure is being resolved by the imager/spectrometer, or, if there is substructure below the instrumental resolving limit, that the strands making up the loop are behaving coherently. Similarly, if a loop with no apparent structuring was observed to be multi-thermal then this could be a clear indicator that the loop consists of unresolved strands or that there us many other structures over a range of temperature along the line of sight. Thus, determining any resolved fundamental spatial scale or the presence of sub-elements within a loop is an important step in addressing how coronal loop (or strand) plasma is possibly being heated.

In an attempt to answer this fundamental question, \citet{schmelz01} constructed multi-thermal DEMs using SoHO/CDS and Yohkoh/SXT. The obtained temperature distributions were found to be inconsistent with isothermal plasma, both along the line of sight and the length of the loop. The advent of TRACE (Transition Region and Coronal Explorer; \citealp{handy99}), yielded more observations on the temperature profiles of coronal loops \citep{schmelz02,winebarger02,winebarger03,cirtain07,schmelz09,tripathi09}. Work by \citet{mulu-moore11} investigate eight active region loops that were previously found to be isothermal \citep{aschwanden05}. Using the cooling-time to loop-lifetime ratio during the rising and decaying phases of the loops, they deduce that observed lifetimes are longer than expected for seven of the loops, suggesting the loops comprise of sub-resolution strands and that many TRACE loops are actually unresolved. \citet{warren02} demonstrate that an impulsively heated loop bundle cools through the TRACE passbands, proposing each strand could appear as a single, long-lived loop with flat 19.5/17.1 nm filter ratios due to the sequential heating of the strands. They argue the model could reproduce observed downflows \citep{winebarger02} and broad DEM distribution along the loops \citep{schmelz01}. Similarly, other models have investigated the many-stranded nature of coronal loops and the impulsive heating through nanoflare events \citep{cargill04, sarkarwalsh08,sarkarwalsh09,taroyan11,price15}.

More recently, high-resolution observations from instruments such as Hi-C and IRIS (Interface Region Imaging Spectrograph; \citealp{depontieu14}) have weighed-in on the discussion of loop widths. For small loops (length $<1$ Mm) \citet{peter13} find widths below 200 km, compared to 1450-2175 km for loops longer than 50 Mm, with no obvious signs of substructure present in the Hi-C data when compared with AIA. From this \citet{peter13} deduce that sub-resolution strands would have to be of the order 15 km wide or smaller for a 1500 km wide loop whose density and temperature vary smoothly across the structure. \citet{brooks13} investigate 91 coronal loops and suggest they are often structured at a scale of several hundred km, ranging between 212 km - 2291 km, with the most frequent occurring FWHM $\approx 640$ km. 

Further work on this by \citet{brooks16} combines IRIS observations and HYDRAD modelling \citep{hydrad} to investigate 108 transition region loops whose FWHM ranges between 266-386~km, arguing that at these spatial scales the structures appear to be composed of monolithic stands rather than composed of multi-stranded bundles.

Similarly, \citet{aschwanden17} combine Monte Carlo simulations of EUV images with the OCCULT-2 loop detection algorithm on the first Hi-C data-set. They find a most frequent distribution of $\approx$550 km for $10^5$ loop width measurements. From this, \citet{aschwanden17} deduce that Hi-C is fully resolving the loop structures. However, when they compare the co-spatial results from AIA they find that AIA can only partially resolve loops $\gtrsim$ 420 km. 

The advancements made by high-resolution measurements of coronal loops, particularly by \citet{brooks16} and \citet{aschwanden17} appear to highlight evidence that current instrumentation is at a stage of resolving individual plasma strands within the corona and hence provides some possible constraint on the heat input required for these features. A summary of coronal loop width measurements can be found in \S5.4.4 of \citet{aschwanden04} along with a table of widths from 52 studies in \citet{aschwanden17}.

This paper undertakes a further examination of loop or strand widths but employs the new data-set obtained from the flight of \hic\ (\S2). Using this unique data-set, we ask if \hic\ is resolving individual coronal strands and what are their spatial scales? To answer this, five regions are investigated from the target active region \ar\ observed with both \hic\ and AIA 171. From these five regions, fourteen cross-section slices are taken which intersect perpendicular to observed coronal strands. Four of these slices are taken from a region of comparative low-emission (\S3.1) compared to ten slices taken from four regions of much higher observed emission (\S3.2). In \S3.3, the widths of the coronal strands are then determined and compared to previous high-resolution findings while the conclusions arising from these results are discussed in \S4.

\section{Hi-C 2.1 Observations and Data Analysis Techniques}
On 29\textsuperscript{th} May 2018 at 18:54 UT, Hi-C was successfully relaunched from the White Sands Missile Range, NM, USA, capturing high-resolution data (2k$\times$2k pixels; $4.4^\prime\times4.4^\prime$ field of view) of target active region \ar. This was the third flight for the Hi-C instrument; the second flight was nominal but the instrument suffered from a shutter malfunction such that no data was captured. For this reason, the mission reported upon here is named \hic. Unlike the first mission that captured Extreme Ultra-Violet (EUV) images in the  narrowband 19.3~nm channel (dominated by Fe XII emission $\approx1.5$ MK), this mission focuses on EUV emission of wavelength 17.2~nm (dominated by Fe IX emission $\approx0.8$ MK), which has a similar temperature response to the AIA 171 passband. \hic\ has a plate scale of 0.129\arc, and captured 78 images with a 2s exposure time and a 4.4s cadence between 18:56 and 19:02 UT. During the \hic\ flight, the instrument experienced a pointing instability resulting in periodic jitter in the dataset. This jitter caused motion blur and lower spatial resolution in approximately half of the data captured. Furthermore, the data were further affected by the shadow of the mesh in the focal plane, reducing the intensity behind the mesh by up to $\approx35\%$. Full details on the \hic\ instrument can be found in \citet{instpaper}.

\begin{figure*}[h!]
\centerline{\includegraphics[width=1.05\textwidth]{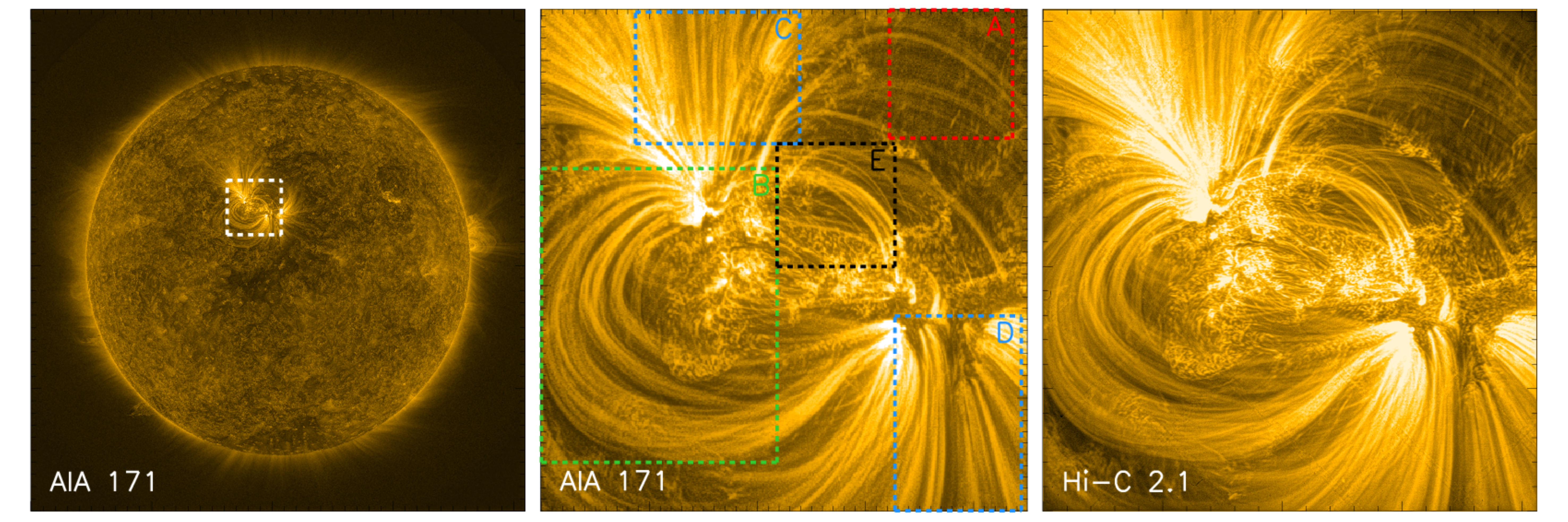}}
\caption{\textit{Left:} Full disk image captured by AIA 171. The dashed-white lines indicate the target active region, AR12712. \textit{Middle:} image for AIA and \textit{right} for \hic. The images have been averaged over the course of $\approx$60 s and sharpened using MGN. The \textit{middle} panel indicates the regions studied in this paper. The low-emission loops (\textit{red}), large loops bundle (green), two open fan regions (blue), and central loops bundle (black) are shown in more detail in Figures~\ref{fig:slices1to4}-\ref{fig:slices13to14}.}
\label{fig:fov}
\end{figure*}

The goal of the analysis outlined here is i) from visual inspection determine a possible range of the structure observed in the Hi-C 2.1 images and ii) to measure the full width half maximum (FWHM) of these detected strands.  This work samples subsections of the \hic\ field of view (FOV), for which the data is time-averaged over a period of $\approx$60~s for both AIA and Hi-C 2.1, taking care to avoid the Hi-C 2.1 exposures impacted by the aforementioned jitter. This helps improve the signal-to-noise ratio, particularly where the emission is low.  Figure~\ref{fig:fov} shows the five regions from which fourteen cross-section slices are taken. Their selection was predicated upon choosing locations where there is possible evidence of substructure or stranding within the loops, whilst taking care to avoid ares in the \hic\ field of view where shadow of the mesh is clearly seen.

To assist in visualizing the finest structures in the Hi-C 2.1 data, we employ an image processing technique. EUV images of the corona span a wide range of features and temperatures, from cooler, low-emission coronal holes, quiet Sun, and filament channels, through to bright, hotter active regions. To account for the dominance of the bright features and reveal low-emission structures often hidden in the data, \citet{morgandruckmuller14} developed the Multi-scale Gaussian Normalisation (MGN) technique for image processing. The method is based on localised normalisation over a range of spatial scales, and thus MGN can reveal the fine detail in the corona and structures in off-limb regions without introducing artifacts or bias. The technique is commonly used in CME and stealth CME detection  \citep{alzate17,hutton17,long18}, but is also used in coronal loop studies \citep{chitta16,long17}.

Due to the way MGN enhances peaks in low-emission plasma, and depresses them in high-emission plasma, the technique is only employed in this work to improve the visual inspection of the AIA and \hic\ data. If the FWHM calculations undertaken in \S3 are done on MGN processed data, it will lead to artificially narrowed(broadened) strands in low(high)-emission plasma. Adapting the method outlined by \cite{pant15} in relation to the first Hi-C mission, two low-frequency passband filters are employed to both the original and MGN sharpened \hic\ datasets to reduce the granular noise present.

\begin{figure}[h]
\centerline{\includegraphics[width=\textwidth]{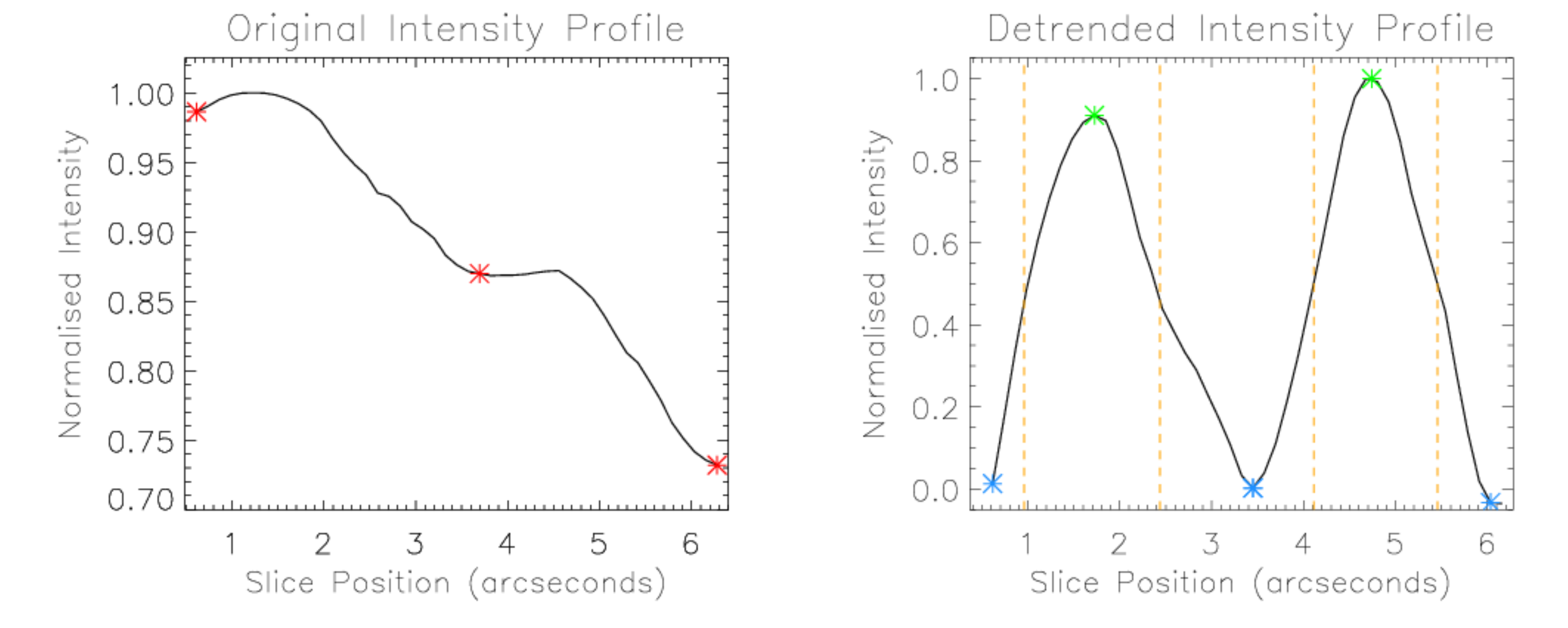}}
\caption{A sub-section of \hic\ intensity slice 10 (\textit{Left}) is shown here to demonstrate the background removal process from the cross-section slices analysed. The red asterisks denote the locations through which we interpolate to obtain the global background. \textit{Right}: The same \hic\ intensity profile with the global background having been deducted. The blue and green asterisks denote the base and maximum values used to determine the height of the peak. The half-maximum is then determined as the midpoint between these two values and is used to determine the FWHM (dashed orange lines).}
\label{fig:fwhm_calc}
\end{figure}

In order to determine the widths of the structures, we  extract the Hi-C 2.1 and AIA intensity along slices inside each sub-region; an example intensity profile is shown in the left panel of Figure~\ref{fig:fwhm_calc}. To further increase the signal-to-noise ratio, the slices in Hi-C 2.1 and AIA are taken to be 3 pixels wide; the intensity is then averaged over these pixels. The slices are taken perpendicular to the strand cross-section in order to obtain accurate measurements of the strand widths. The slice locations and averaged intensities in AIA and Hi-C 2.1 are given in Figures~\ref{fig:slices1to4}-\ref{fig:slices13to14} in the first and second columns, respectively.  After finding the average intensity along each slice, the global trend is removed by finding all the local minima of a slice (shown in Figure~\ref{fig:fwhm_calc} as red asterisks), and interpolating through these values. The resulting trend is then subtracted from the intensity profile of the slice, leaving behind the variations, i.e. the coronal strands, seen in the right panel of Figure~\ref{fig:fwhm_calc} and the third columns of Figures~\ref{fig:slices1to4}-\ref{fig:slices13to14}. The base locations of the isolated coronal strands are determined as the inflection points, and the maximum value between the two inflection points is taken as the maxima of each strand. From these two values, the half-maximum value is determined and their locations are used to determine the FWHM of each structure analysed (orange dashed lines in Figure~\ref{fig:fwhm_calc}). These FWHM values are used as a possible determination of the coronal strand widths, which are then compared to previous high-resolution studies.

The uncertainty in the intensity for the cross-sectional slices is determined by $\Delta I = \sqrt{I}$. If the slice under consideration is from a region where the intensity is low, $\Delta I$ will correspond to a larger proportion of the intensity than a slice from a region of greater intensity. In some instances, such as slice 10 (Figure~\ref{fig:slices8to10}), the magnitude of $\Delta I$ appears to increase once the background subtraction has occurred (variation slice 10). This is merely a consequence of the background emission being large relative to the local intensity of the coronal strands itself. Re-normalising the background subtracted slices has the effect of focusing in on the structures themselves, which thus means that $\Delta I$ will appear larger.

\section{Results}
In this study cross-sectional profiles are taken from structures observed within five regions, whose positions in the AIA and \hic\ FOV are indicated in Figure~\ref{fig:fov}. Table~\ref{table:emission} indicates the average emission for all five regions investigated during the $\approx 60 s$ for which the data is time-averaged. As is seen in Table~\ref{table:emission}, the average emission of Region A is at least an order of magnitude lower than the other locations investigated within \ar. For this reason, the study is split into two parts in what we term in this paper as low-emission loops (Region A in Figure~\ref{fig:fov}), and four high-emission loop regions (Regions B-E in Figure~\ref{fig:fov}). The low-emission loops are shown in more detail in Figure~\ref{fig:slices1to4}, whilst the high-emission regions include a selection of loops; large loops (Figure~\ref{fig:slices5to7}), open fan loop regions (Figures ~\ref{fig:slices8to10}~\& \ref{fig:slices11to12}), and some small loops close to the centre of the active region (Figure~\ref{fig:slices13to14}).
\begin{table}
\centering
\caption{Average emission for all pixels in the regions studied over the $\approx60$ s time-averaged data.}
\begin{tabular}{ |c|c|c|c| }
\hline
Region in Figure~\ref{fig:fov} & Loop Type & Mean Emission \hic & Mean Emission AIA  \\
 &  & (DN/pixel) & (DN/pixel) \\
\hline
A & Low-Emission Loops & 1229.13 & 211.550 \\
B & Large Loops Bundle & 44892.5 & 3792.84  \\
C & Northern Open Fan Loops & 79550.3 & 6939.30  \\
D & Southern Open Fan Loops & 55688.6 & 4983.92  \\
E & Central Loops Bundle & 46270.3 & 3720.91  \\
\hline
\end{tabular}
\label{table:emission}
\end{table}
\begin{figure}[h!]
\centerline{\includegraphics[width=0.8\textwidth]{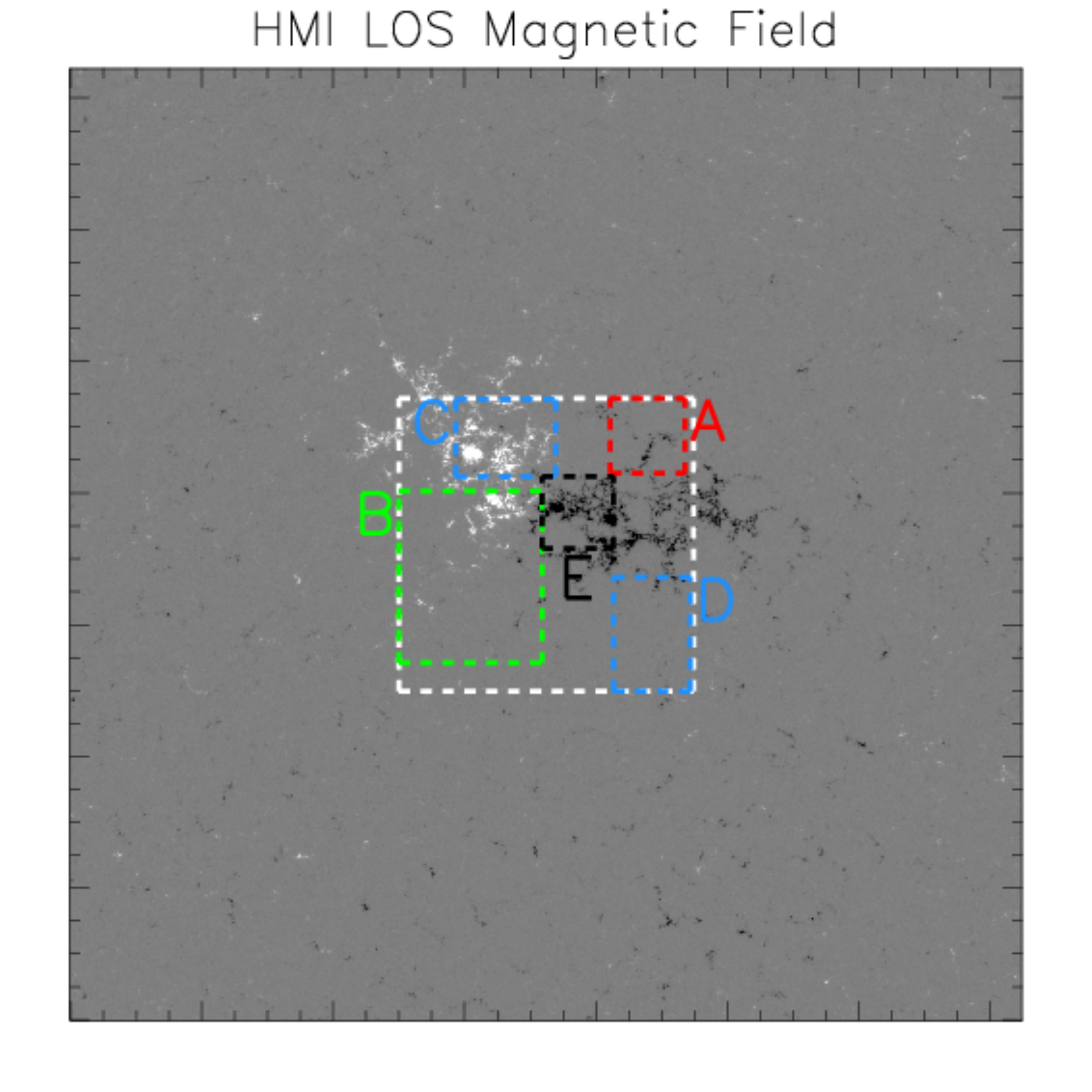}}
\caption{HMI Line-of-sight (LOS) Magnetic Field for the \hic\ FOV (white) and the surrounding area. The five regions (A-E) examined in the study are indicated in the same manner as Figure~\ref{fig:fov}.}
\label{fig:hmi_fov}
\end{figure}

Figure~\ref{fig:hmi_fov} shows the the Helioseismic and Magnetic Imager (HMI) line-of-sight (LOS) magnetic field for the \hic\ FOV and the surrounding area. The snapshot shown here is taken at 18:56:15 UT, which corresponds to the $\approx60$~s period under examination with \hic\ and AIA. From this it can be seen that AR 12712 resides above a diffuse bipolar region.

The closed magnetic loops under investigation in regions B and E very clearly have their foot-points rooted in the areas of opposite polarity, likely crossing over the active region's polarity inversion line. The open fan loops observed in regions C and D originate from areas of opposite polarity. The low-emission strands observed in region A have one footpoint in the area of positive polarity but their negative polarity footpoint lies outside of the \hic\ field of view. However, these low emission strands are still an integral part of the overall active region itself.

The strands in region A are low-density, and subsequently low-emission. However, due to their location and ideal viewing angle placing them away from the core of the active region, these strands can be observed in a more isolated manner, helping us to determine their widths. The coronal magnetic field is often considered to be force-free in 3D simulations (\citealp{aschwanden19} and references therein), which implies that current density scales with the magnetic field (e.g. Figure~1 in \citealp{gudiksen02}). One may expect where current density is larger for the heating to be greater, and subsequently the emission to be higher. For these reasons, the study is separated into low-emission (\S3.1) and high-emission (\S3.2) regions.

\subsection{Low-Emission Loops} 
\begin{sidewaysfigure}[hbtp]
\centerline{\includegraphics[width=\textwidth]{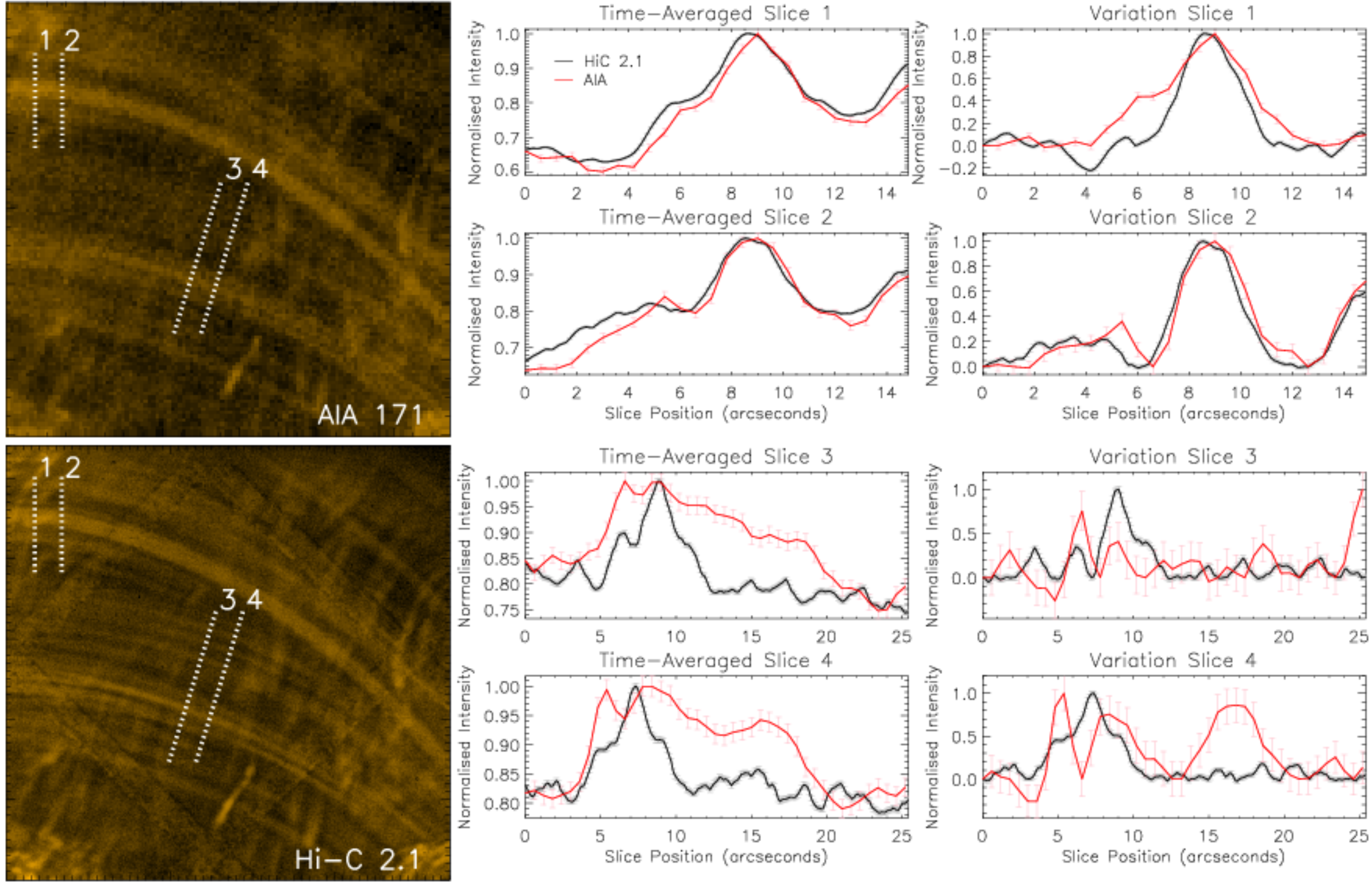}}
\caption{The two images in the left column indicate the location of the low-emission loops for AIA (\textit{top}) and for \hic~(\textit{bottom}). The intensity slices are numbered 1 through 4 indicating the cross-sections under consideration. The middle column shows the time-averaged plots of intensity along the cross-sectional slices for AIA (\textit{red}) and \hic~(\textit{black}). Here, slice position from left-to-right corresponds to south-to-north orientation in the images. The right column shows the detrended, time-averaged variations of each slice, i.e. the slices with the global trends having been subtracted. The error bars indicate the uncertainty in the intensity, which is defined as $\Delta I = \sqrt{I}$.}
\label{fig:slices1to4}
\end{sidewaysfigure}
\begin{sidewaysfigure}[hbtp]
\centerline{\includegraphics[width=\textwidth]{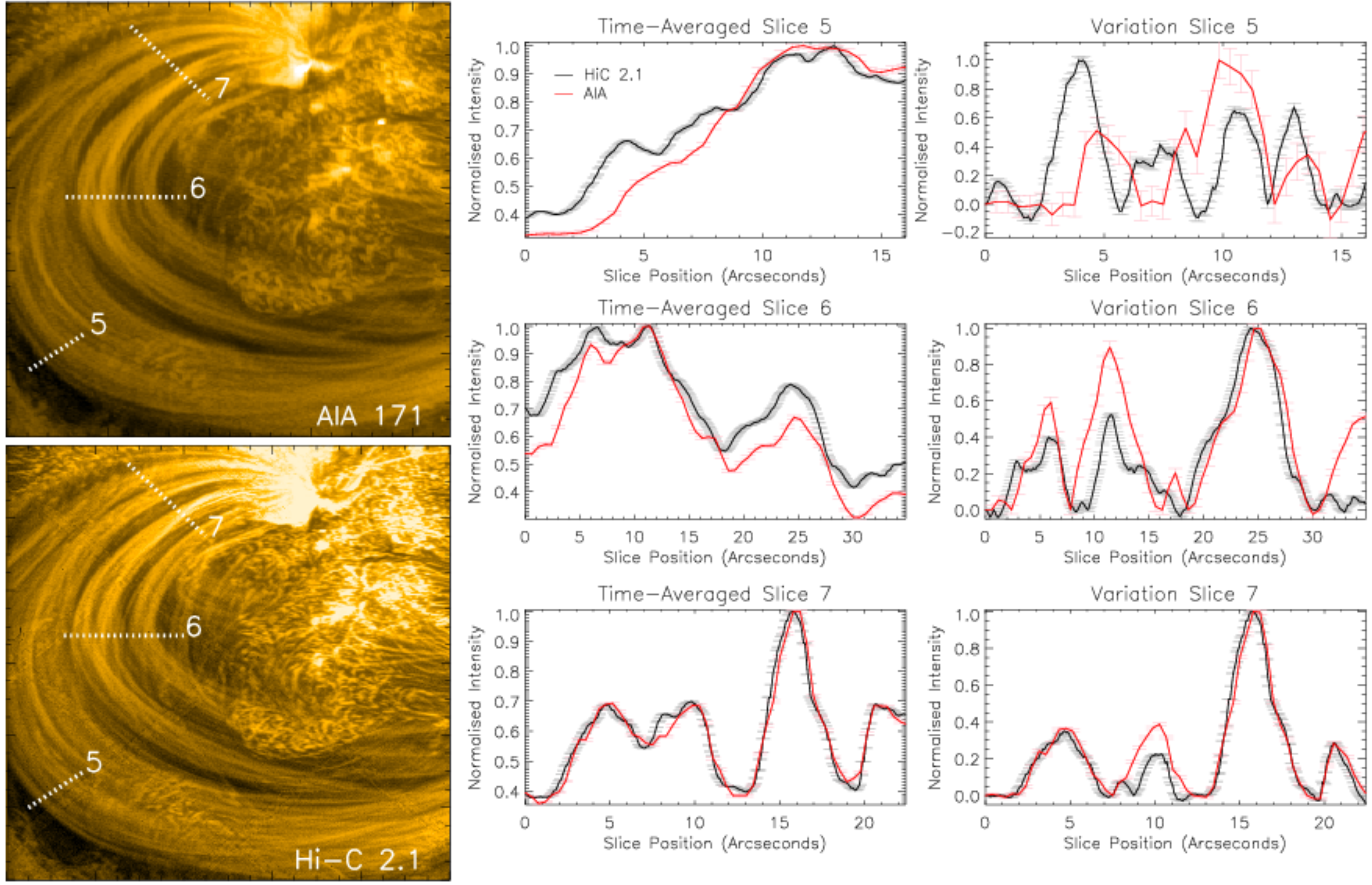}}
\caption{The same layout and approach as in} Figure~\ref{fig:slices1to4} but for the large loops bundle in the south west quadrant of the \hic\ FOV. Note: slice position left-to-right in the intensity plots corresponds to west-to-east orientation in the images. The error bars indicate the uncertainty in the intensity, which is defined as $\Delta I = \sqrt{I}$.
\label{fig:slices5to7}
\end{sidewaysfigure}

Given the way that MGN normalises a range of emission, figure~\ref{fig:slices1to4} indicates visually the fine-scale structure that is present even in regions which may initially appear to be beneath the detection threshold. The top left panel shows the MGN sharpened AIA time-averaged data, and the bottom left panel shows the corresponding MGN sharpened \hic~ image. From these two images, it can be seen that AIA does not have the resolution to differentiate some of the lower density, low-emission strands above the background corona and instrumental noise. \hic\ on the other hand, performs much better with several, low-density strands being observably distinguishable.

Four data slices are taken across Region A (numbered 1-4 in Figure~\ref{fig:slices1to4}) and the normalised emission intensity along each slice is compared for \hic\ and AIA. Each intensity profile is plotted from south to north. Slices 1 and 2 show good agreement between \hic\ and AIA, particularly with the broader structure centred around 9\arc. However, there are signs of the structure consisting of more than a single strand in \hic\ due to the irregular shape. South of this in slice 2 there is a single strand in the AIA data ($2\arc - 6.5\arc$) but in \hic\ there is evidence of three coronal strands, which trace the same envelope resolved by AIA. These \hic\ structures have FWHM between $345 - 396$ km, which is beneath the width of a single AIA pixel.

The difference between the two instruments becomes more apparent in slices 3 and 4. In the MGN sharpened images, it can be seen in the southernmost part of the slices that there are three distinct strands for \hic\ and one to two strands in the corresponding AIA envelope. However, in the time-averaged slices for the non-MGN processed \hic\ data, only two peaks can be seen in slice 3 (between 5\arc - 12\arc) and one peak in slice 4 (3\arc - 11\arc). This arises due to the way in which MGN normalizes and enhances over a range of spatial scales and thus cannot be employed for direct data analysis in this study.

Further north of this large envelope ($12\arc - 25\arc$), it can be seen that there is plasma detected by AIA, but this is noisy with no structure being resolved. This can be seen in the time-averaged slices as the normalised intensity value is between $\approx0.85-0.95$ but no individual peaks can be identified that are above the error bars. For \hic\ however, there are multiple strands resolved in this region, with five strands resolved between $14\arc-25\arc$ in slice 3 , and eight strands between $11\arc-24\arc$ in slice 4.

\subsection{High-Emission Measure Loops}
\begin{sidewaysfigure}[hbtp]
\centerline{\includegraphics[width=\textwidth]{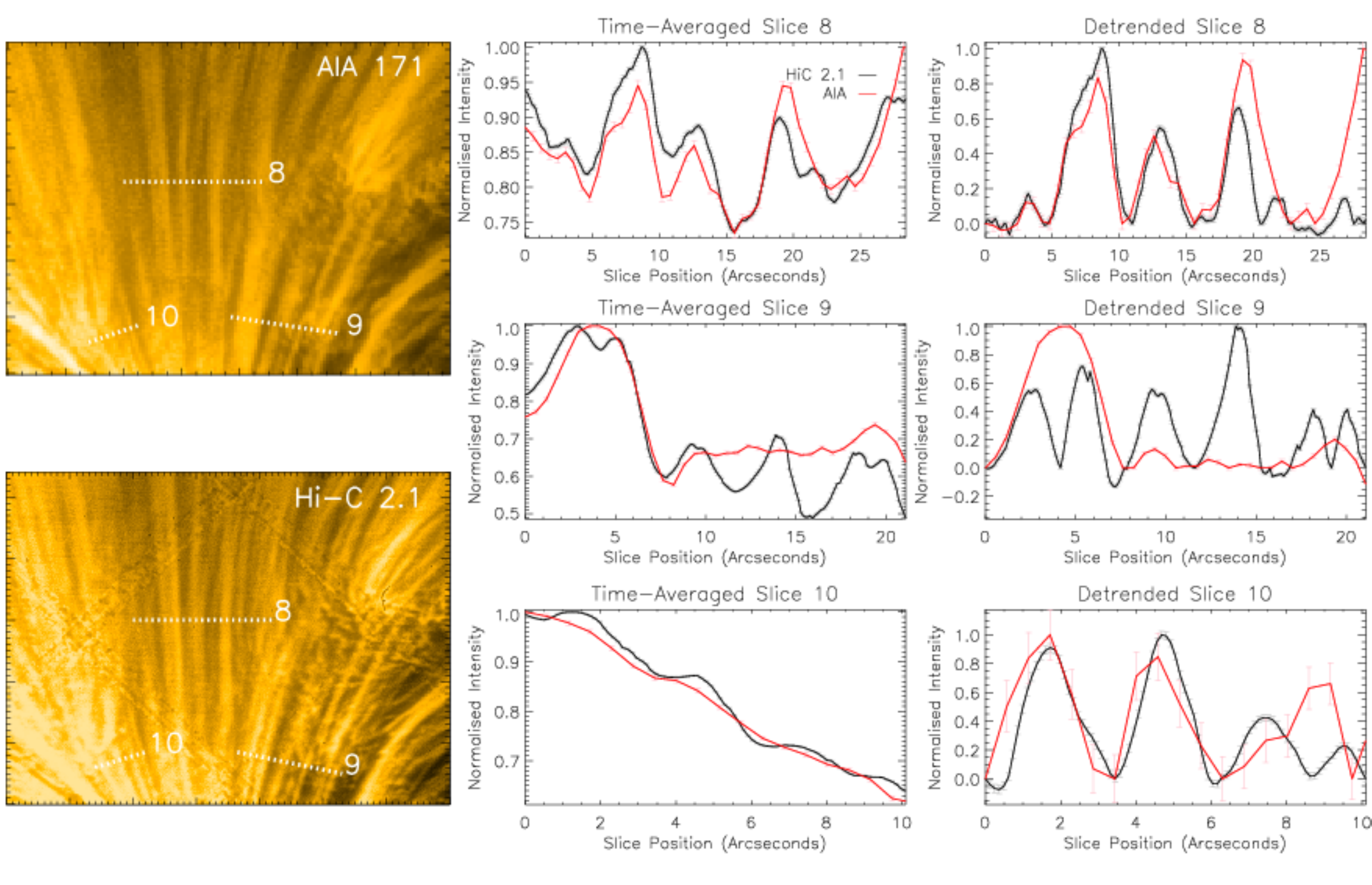}}
\caption{The same layout and approach as in} Figure~\ref{fig:slices1to4} but for the north open fan loops in the northwest quadrant of \hic\ FOV. Note: slice position left-to-right in the intensity plots corresponds to west-to-east orientation in the images. The error bars indicate the uncertainty in the intensity, which is defined as $\Delta I = \sqrt{I}$.
\label{fig:slices8to10}
\end{sidewaysfigure}
\begin{sidewaysfigure}[hbtp]
\centerline{\includegraphics[width=\textwidth]{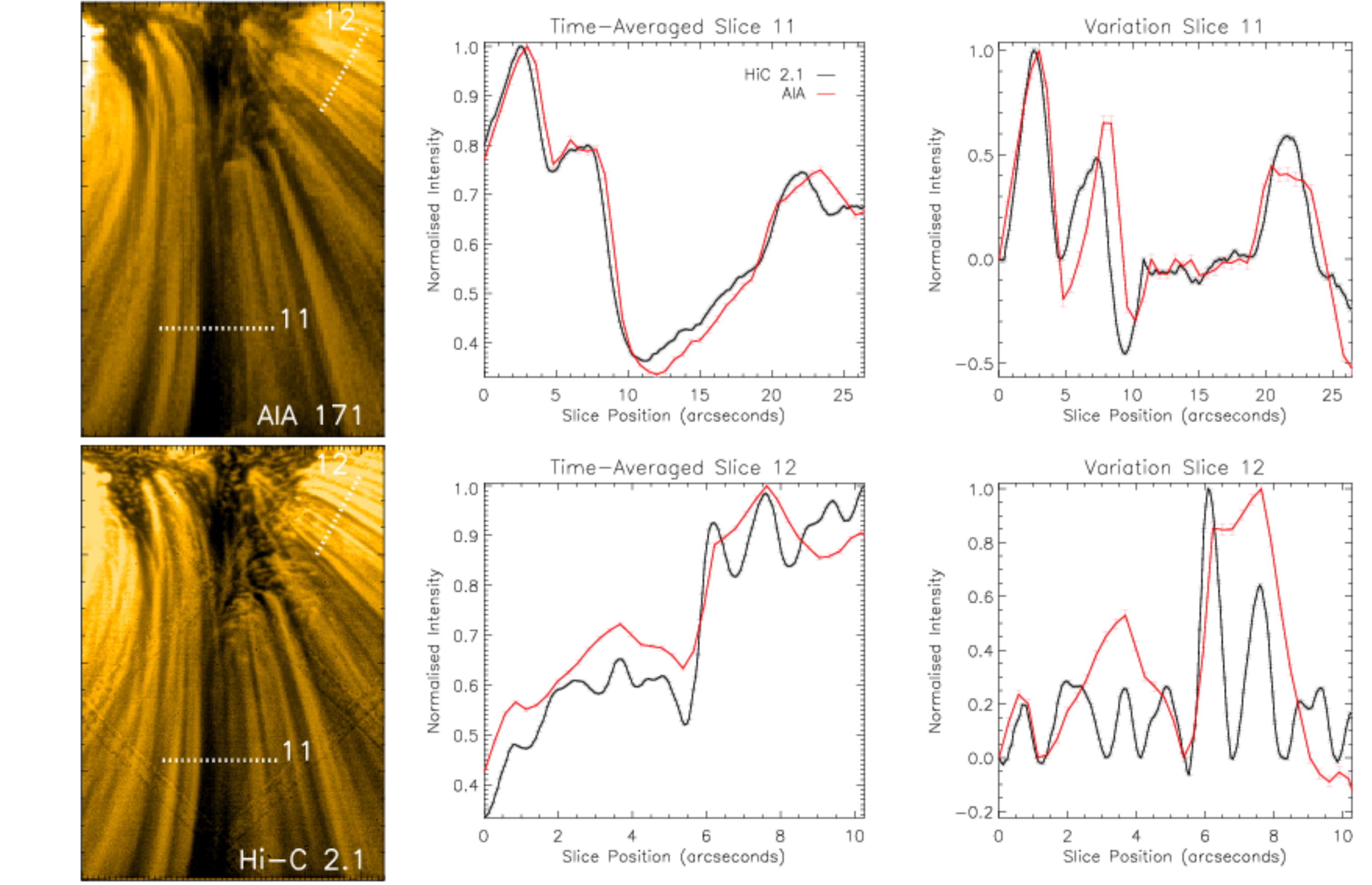}}
\caption{The same layout and approach as in} Figure~\ref{fig:slices1to4} but for the south open fan loops in the southeast quadrant of the \hic\ FOV. Note: slice position left-to-right in the intensity plots corresponds to west-to-east orientation in the images. The error bars indicate the uncertainty in the intensity, which is defined as $\Delta I = \sqrt{I}$.
\label{fig:slices11to12}
\end{sidewaysfigure}
\begin{sidewaysfigure}[hbtp]
\centerline{\includegraphics[width=\textwidth]{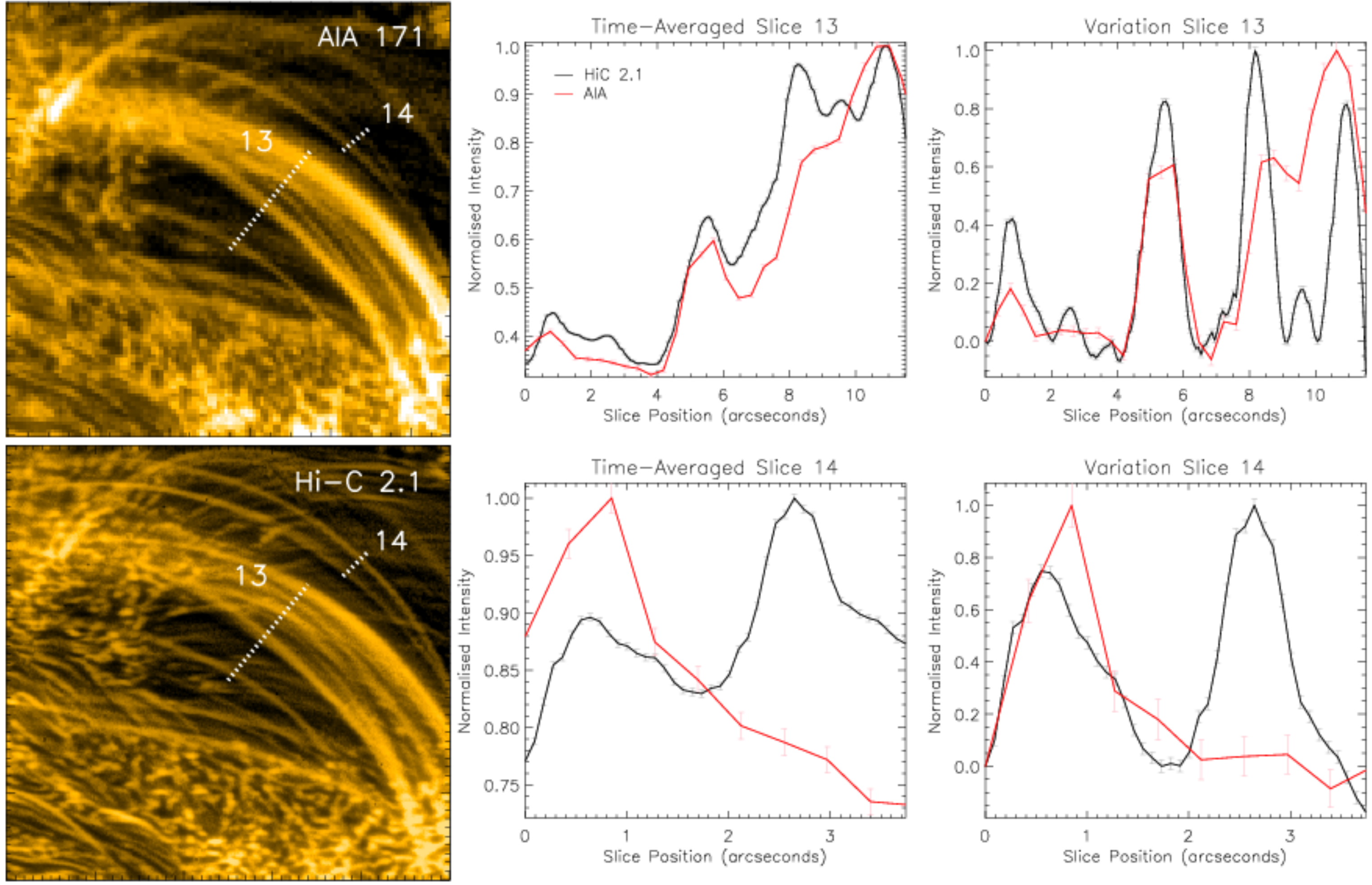}}
\caption{The same layout and approach as in} Figure~\ref{fig:slices1to4} but for the central loops bundle between the central moss and low-emission loops. Note: slice position left-to-right in the intensity plots corresponds to south-to-north orientation in the images. The error bars indicate the uncertainty in the intensity, which is defined as $\Delta I = \sqrt{I}$.
\label{fig:slices13to14}
\end{sidewaysfigure}

In this subsection results from investigations of ten cross-sectional slices in the high-emission loop regions (B-E in Figure~\ref{fig:fov}) are presented in the same manner as the low-emission loops discussed in \S 3.1. Again, the normalised intensity profiles are plotted but left-to-right orientation now corresponds to west-to-east in the respective AIA and \hic\ images (Figures~\ref{fig:slices5to7} - \ref{fig:slices13to14}).

The cross-sectional slices typically show similar intensity profiles for \hic\ and AIA, with many structures being nearly identical. In particular, slices 7 (Figure \ref{fig:slices5to7}), 8 \& 10 (Figure \ref{fig:slices8to10}), and 11 (Figure \ref{fig:slices11to12}) display strikingly similar overall profiles in \hic\ and AIA. The only appreciable differences seen in slices 7 (7\arc-12\arc), 8 (17\arc-23\arc), and 10 (6\arc-10\arc) occur where single AIA strands correspond to two \hic\ strands, which have widths of 562 km \& 1146 km in slice 7, 1333 km \& 923 km in slice 8, and 985 km \& 556 km in slice 10, placing them approximately between one and three AIA pixel widths.

Similarly, there are also examples where although \hic\ and AIA observe the same general structure, \hic\ potentially resolves more coronal structures along the length of the cross-sections, such as is seen in slices 5 \& 6 (Figure \ref{fig:slices5to7}), and slice 13 (Figure \ref{fig:slices13to14}). In these three slices there is reduced commonality between the two instruments as the variations are increased compared to the nearly identical cross-sections seen in slices 7, 8, 10, and 11. Focusing on slice 5, the time-averaged intensity plots of both instruments appear to show agreement; however, the corresponding detrended profile (variation plot 5 in Figure~\ref{fig:slices5to7}) reveals important differences. This can most notably be seen with the AIA structure centred at 5\arc\ spanning between two \hic\ strands. Further along slice 5, the AIA structure is double peaked, with the corresponding emission coming from two distinct structures in \hic. Other examples can be seen in slices 6 (0\arc-19\arc) and 13 (1.5\arc-4\arc\ \& 7\arc-13\arc).

It is slices 9, 12, and 14 that the difference in resolution, and subsequently resolving power of the two instruments is most notable for the high-emission regions. Between slice position 0\arc-8\arc in slice 9 (Figure \ref{fig:slices8to10}) there is a single, large structure in AIA, which can be seen as two peaks in the time-averaged \hic\ data. This double peak is evident in the corresponding MGN \hic\ image, whilst in the MGN AIA image the structure still appears monolithic. 

Closer inspection of the two \hic\ structures (0\arc-7\arc) in the variation plot reveals that the `monolithic' AIA structure could actually be composed of four strands. This is because, whilst the detrended profiles of the \hic\ structures are large, the peaks of the two structures are actually comprised of two small-peaks, which are fully-resolved and above the error bars. 

The next AIA structure in slice 9 also appears as a broad, mostly unresolved single feature being detected at position 19\arc. However, in the corresponding \hic\ data there are two strands. Similarly, for the rest of the slice, \hic\ detects two more strands; both indicating signs of possible substructure with faint double-peaks like the two aforementioned \hic\ structures ($0\arc-7\arc$). This highlights that there is evidence for further substructuring beyond anything that \hic\ can observe.

In slice 12 (Figure \ref{fig:slices11to12}) we see one of the clearest examples where the increased resolving power of \hic\ reveals strands which AIA does not distinguish. The two large AIA features between 1\arc-5\arc\ and 5\arc-9\arc each correspond to three \hic\ strands, which have a mean width of 579 km. In Figure \ref{fig:slices13to14}, slice 14 samples the cross-section of two strands which are relatively isolated against the underlying moss region. The southern-most strand ($\approx0\arc-1.6\arc$) shows good agreement between \hic\ and AIA in the variation plot, though again there is a non-smooth, irregular distribution in the \hic\ data hinting at unresolved coronal strands. The northern-most strand ($\approx1.9\arc-3.7\arc$) is well defined in \hic\ with no obvious signs of substructure. However, in AIA there is no structure/peak at this location. The corresponding MGN sharpened AIA image reveals that the strand fades in-and-out of detection along its length, indicating this strand is at the detection threshold of AIA. This could mean that either this is a low-emission strand, or the strand has a very narrow or precise temperature which is very close to the peak emission temperature of \hic.

\subsection{Strand Widths}
\begin{figure}[t]
\centerline{\includegraphics[width=0.8\textwidth]{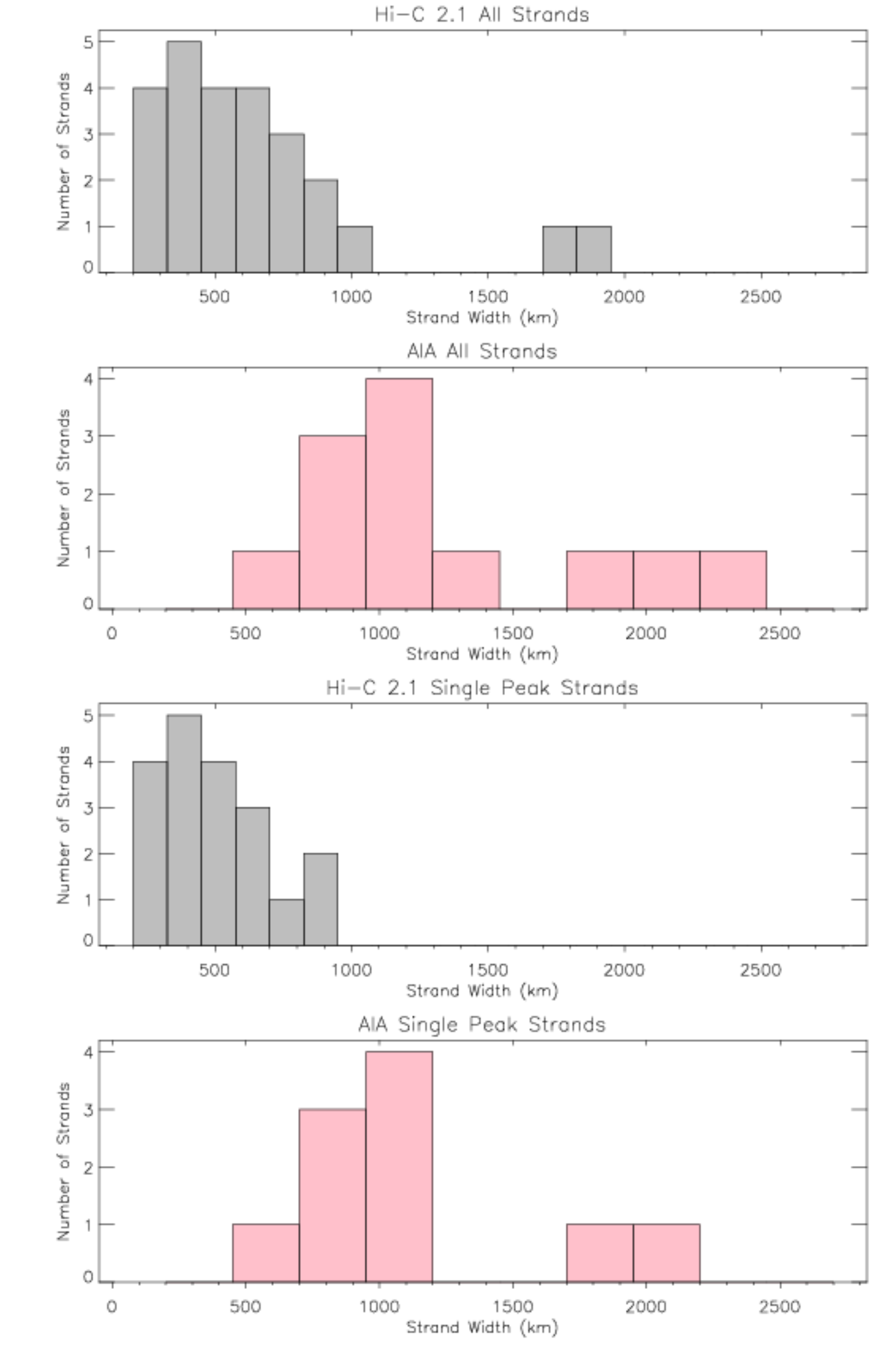}}
\caption{This figure collates the strand widths measured for the low-emission region. The top two panels display occurrence frequency plots of strand width against number of strands observed for all the strands measured using \hic\ and AIA. The bottom two panels show the same occurrence frequency plots but now for 19 single-peak strands in \hic\ and 10 single-peak strands in AIA. The \hic\ widths are shown in 125 km bins, whereas the AIA bin width is 250 km.}
\label{fig:fwhm_le}
\end{figure}
\begin{figure}[t]
\centerline{\includegraphics[width=0.8\textwidth]{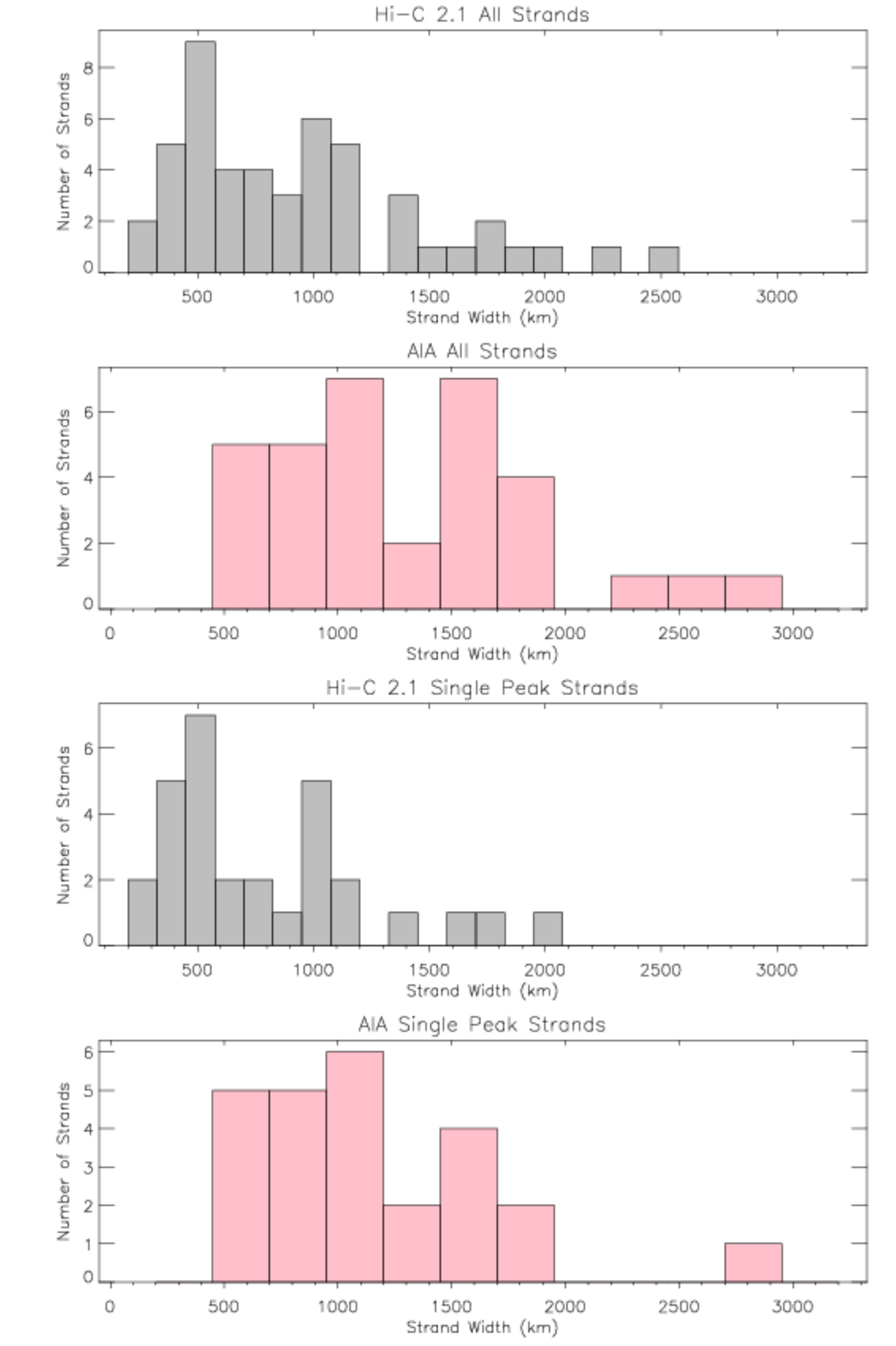}}
\caption{This figure collates the strand widths measured for the high-emission regions. The top two panels display occurrence frequency plots of strand width against number of strands observed for all the strands measured using \hic\ and AIA. The bottom two panels show the same occurrence frequency plots but now for 30 single-peak strands in \hic\ and 25 single-peak strands in AIA. The \hic\ widths are shown in 125 km bins, whereas the AIA bin width is 250 km.}
\label{fig:fwhm_ar}
\end{figure}
A total of 25 and 49 strand widths are measured in the low-emission and high-emission regions, respectively. The appendix contains two tables that index the widths and locations of all the strands measured by \hic\ and AIA for the low-emission (Table~\ref{table:LEloops}) and high-emission regions (Table~\ref{table:ARloops}). 

However, as can be seen in slices 3 and 12, some of the structures are double-peaked. Consequently, these result from either the presence of strands beneath the resolving power of \hic\ or due to other coronal plasma somewhere else along the integrated line-of-sight due to the plasma being optically thin. Measuring the widths of these structures could lead to an artificially broadened distribution. Subsequently, analysis is undertaken with these double-peaked structures discarded from this survey, hence leaving 19 low-emission strands and 30 high-emission strands for width analysis. Thus, the widths are collated into occurrence frequency plots in Figure \ref{fig:fwhm_le} for the low-emission strands and Figure \ref{fig:fwhm_ar} for the high-emission strands.

The \hic\ occurrence frequency plot for the low-emission strand widths (Figure \ref{fig:fwhm_le}) reveals there are low-density, low-emission strands on both ends of the spatial scale, with structures as small as $\approx200$ km, and as large as $1700-1950$ km. However, these broader structures have irregular intensity profiles (as mentioned in \S 3.1). Consequently, when only single-peaked structures are considered, the broader strands are between $825-950$ km, which is narrower than the most likely strand width measured with AIA ($950-1200$ km). For \hic, the most likely strand width of the low-emission structures is $325-450$ km, placing the structures at approximately the same width as an AIA pixel or smaller.

Figure~\ref{fig:fwhm_ar} shows a similar distribution of strand widths for the four high-emission regions as is seen with the low-emission region. Again, narrow strands $\approx200$ km wide are detected by \hic\ though these are not as prevalent as in the low-emission region despite a larger sample of structures being investigated. Additionally, broader strands with widths $>1000$ km are observed in the high-emission regions, with one strand in slice 11 ($\approx19\arc-24.6\arc$) exceeding a width of 2000 km. The most likely strand width for single-peaked structures in the four high-emission regions is between $450-575$ km. Considering the smallest partially resolved structure by AIA ($466$ km) is larger than the average low-emission strand width resolved by \hic\ ($\approx388$ km), it could be argued that these results, alongside those from the maiden flight of Hi-C, provide compelling evidence for satellite-borne instrumentation with the resolving power of at least Hi-C.

For AIA, the most likely strand width of the low-emission and high-emission regions are the same ($\approx1075$ km) but for \hic\ the low-emission strands are approximately 125 km narrower than those resolved in the high-emission region ($\approx388$ km vs $\approx513$ km). It is worth noting that there is a greater proportion of the low-emission strands with widths between $200-325$ km compared to the high-emission strands. Conversely, the high-emission strands often exhibit widths $\gtrapprox1000$ km, agreeing with results from \citet{peter13}, whilst the low-emission strand widths do not exceed $950$ km in this study.

\section{Concluding Remarks}
Continuing on from the success of Hi-C, \hic\ reveals structures throughout and around its targeted active region (\ar) in the 17.2 nm line that cannot be resolved by SDO/AIA. The work outlined here investigates five regions from the \hic\ FOV, one region of what could be considered low-emission and four regions of significantly increased emission (Figure~\ref{fig:fov}).

In regard to Region A, although  it could be argued that there are hints of faint structures that could be observed in the AIA FOV, it is found that with \hic's superior resolving power, this region is in fact filled with numerous low emitting, and hence low-density strands. In AIA, these strands often appear as granular noise. A contributing factor to this is that these low-emission strands are observed by \hic\ to be of only $\approx388$ km in width, beneath the scale of a single AIA pixel.

In contrast, Regions B-E have higher emission structures with an average \hic\ strand width of $\approx513$ km, placing them in line with previous width measurements made using Hi-C \citep{brooks13,aschwanden17}. As discussed in \citet{peter13} with regards to miniature loops, the \hic\ data reveals strand widths as small as 200 km in both the low-emission and high-emission regions, though they appear to be more numerous in the low-emission area of the active region.

An intriguing result is outlined in the analysis of slice 14, which samples the cross-section of the closed active region loops near the centre of the \hic\ FOV.  In the MGN sharpened AIA image (Figure~\ref{fig:slices13to14}), the northern-most strand sampled in slice 14 appears to fade in-and-out of AIA detection along its length. This suggests that the plasma contained within this structure is at the sensitivity limit of AIA, even when advanced image processing techniques like MGN are employed. Either this is a low-emission strand amongst the high-emission structures, or this may simply be due to the fact that AIA and \hic\ observe plasma at slightly different temperatures (17.1nm and 17.2nm emission).

The evidence for low-emission strands, which are very difficult to observe with AIA but much better resolved and width determined by \hic, strongly indicates that plasma threads with low-density but coronal temperature material are prevalent throughout the corona. This may be a strong indicator of previously unresolved, but background heated corona that \hic\ is beginning to provide evidence for. However, even with the enhanced spatial resolution of \hic, it still appears that there are structures which could not be fully resolved. Although this could be due to projection effects of the optically thin plasma viewed along its line-of-sight, another possible scenario is that there are coronal strands with structural widths below even the resolving power of \hic. 

Slices 9 \& 12 are good examples of this but there are hints of substructure above the observational error throughout all the slices examined. Thus, it may be possible for an instrument of greater resolving power to discriminate between these features. Note that by combining both FFT and Gaussian width analysis methods, \citet{instpaper} conclude that the \hic\ resolution is between $0.3-0.47\arc$ ($\approx220-340$~km) in images that are not affected by motion blur. The most-likely strand widths obtained for the low-emission strands ($\approx388$~km) are above the resolution limit of \hic, indicating that at this increased spatial resolution we may be beginning to have the opportunity resolve a fundamental width of individual coronal strands. This result agrees with the results from \citet{aschwanden17} though we note that there is building evidence that there are strands with widths beneath the \hic\ resolution.

Additionally, it may be argued that spatial structuring is only one part of the data required to address this question of basic plasma stranding as the spread of observed temperature of these features is also critical. If the 17.2 nm \hic\ structures that are observed to be beneath the AIA resolution limit could also be observed in other passbands with similar spatial resolution to \hic\ and those observations then resulted in a broad temperature distribution of many strand-like features, then this could be strong evidence for multi-thermal, many stranded models being the best way tackle coronal heating.

Future work will further examine the double-peaked structures by fitting appropriate Gaussian profiles in order to estimate the widths of the possible sub-resolution strands, as well as consider the examination of the extrapolated magnetic field structure associated with low-emission strands alongside a comparison of the \hic\ observations with specific active region modelling \citep{peter19}.

\begin{acknowledgements}
We acknowledge the High-resolution Coronal Imager (Hi-C 2.1) instrument team for making the second re-flight data available under NASA Heliophysics Technology and Instrument Development for Science (HTIDS) Low Cost Access to Space (LCAS) program (proposal HTIDS17\_2-0033). MSFC/NASA led the mission with partners including the Smithsonian Astrophysical Observatory, the University of Central Lancashire, and Lockheed Martin Solar and Astrophysics Laboratory.  Hi-C 2.1 was launched out of the White Sands Missile Range on 2018 May 29. S.K.T. gratefully acknowledges support by NASA contracts NNG09FA40C (IRIS), and NNM07AA01C (Hinode). The work of DHB was performed under contract to the Naval Research Laboratory and funded by the NASA Hinode program.
\end{acknowledgements}


\appendix
In this appendix we present Tables~\ref{table:LEloops} \& \ref{table:ARloops}. These show all the strands where FWHM calculations were possible. The widths shown in bold with an asterisks denote strands which display obvious signs of sub-resolution and/or overlapping strands, and are thus omitted from final statistical analysis on the widths.
\begin{table}[htb]
  \centering
  \caption{FWHM for the Investigated Low-Emission Loops}
  \label{table:LEloops}
  \begin{tabular}{*8c}
    \toprule
    \multicolumn{4}{c}{\hic}  & \multicolumn{4}{c}{AIA 171\AA}  \\ \cmidrule(r){1-4} \cmidrule(l){5-8}
    Slice \# & Start Position (\arc) & End Position (\arc) & FWHM (km)         &   Slice \# & Start Position (\arc) & End Position (\arc) & FWHM (km)  \\ \cmidrule(r){1-4} \cmidrule(l){5-8}
    1 & 0.000 & 1.968 & 697.5 & & & & \\
    1 & 1.968 & 3.150 & 473.6 & & & & \\
    1 & 4.987 & 5.906 & 381.2 & 1 & 4.200 & 13.20 & \textbf{2383.8*} \\
    1 & 5.906 & 11.28 & \textbf{1819.4*} & & & & \\
    1 & 12.20 & 13.51 & 409.8 & & & & \\
    2 & 0.000 & 1.050 & 226.6 & & & & \\
    2 & 1.050 & 2.625 & 345.2 & 2 & 1.800 & 6.600 & \textbf{1373.4*} \\
    2 & 2.625 & 3.937 & 359.4 & & & & \\    
    2 & 4.200 & 5.643 & 395.5 & 2 & 6.600 & 12.60 & 2041.6 \\
    2 & 6.037 & 11.81 & \textbf{1859.0*} & & & & \\
    3 & 1.806 & 4.644 & 733.9 & 3 & 0.600 & 4.200 & 838.7 \\
    3 & 5.031 & 7.224 & \textbf{822.0*} & 3 & 5.400 & 7.800 & 853.2 \\
    3 & 7.224 & 10.57 & 928.9 & 3 & 7.800 & 10.80 & 1062.1 \\    
    3 & 14.06 & 15.73 & 552.0 & 3 & 15.00 & 17.40 & 642.6 \\
    3 & 16.12 & 18.31 & 590.9 & 3 & 17.40 & 21.00 & 1130.1 \\    
    3 & 18.83 & 21.02 & \textbf{972.0*} & 3 & 21.00 & 23.40 & 925.7 \\
    3 & 21.02 & 22.96 & \textbf{620.5*} & & & & \\
    4 & 11.61 & 12.25 & 215.7 & 4 & 4.200 & 6.600 & 951.2 \\
    4 & 12.65 & 13.67 & 664.6 & & & & \\    
    4 & 14.06 & 14.96 & 318.0 & 4 & 6.600 & 10.80 & 1829.0 \\
    4 & 14.96 & 16.12 & 527.4 & & & & \\    
    4 & 16.12 & 16.89 & 216.8 & 4 & 13.20 & 21.00 & 2701.2 \\
    4 & 17.93 & 19.35 & 466.7 & & & & \\
    4 & 19.73 & 21.67 & 863.7 & 4 & 21.60 & 24.60 & 959.9 \\
    4 & 21.67 & 23.09 & \textbf{717.4*} & & & & \\
    \bottomrule
  \end{tabular}
\end{table}
\begin{table}[htb]
  \centering
  \caption{FWHM for the Investigated High-Emission Loops}
  \label{table:ARloops}
  \begin{tabular}{*8c}
    \toprule
    \multicolumn{4}{c}{\hic}  & \multicolumn{4}{c}{AIA 171\AA}  \\ \cmidrule(r){1-4} \cmidrule(l){5-8}
    Slice \# & Start Position (\arc) & End Position (\arc) & FWHM (km)         &   Slice \# & Start Position (\arc) & End Position (\arc) & FWHM (km)  \\ \cmidrule(r){1-4} \cmidrule(l){5-8}
    5 & 0.000 & 1.712 & 528.5  & & & & \\
    5 & 1.913 & 5.640 & \textbf{1346.6*} & 5 & 3.748 & 6.559 & 1547.6 \\
    5 & 9.267 & 11.88 & \textbf{1190.5*} & 5 & 8.901 & 12.18 & 1513.0 \\
    5 & 11.88 & 14.20 & 871.2  & 5 & 12.18 & 14.52 & 1137.4 \\
    5 & 14.40 & 15.31 & 202.2  & & & & \\
    6 & 4.515 & 8.256 & 1046.0 & 6 & 0.600 & 2.400 & 895.2 \\
    6 & 8.256 & 9.417 & 355.8  & 6 & 2.400 & 7.800 & \textbf{1781.3*} \\
    6 & 9.417 & 13.15 & 1047.7 & 6 & 7.800 & 16.20 & 2840.0 \\
    6 & 30.18 & 32.89 & 1107.1 & 6 & 16.20 & 18.60 & 893.8 \\
    7 & 1.783 & 7.220 & \textbf{2304.5*} & 7 & 1.975 & 7.506 & \textbf{2279.2*} \\
    7 & 7.220 & 8.749 & 562.0  & 7 & 7.506 & 13.03 & 1577.4 \\
    7 & 8.749 & 11.38 & 1146.1 & 7 & 13.03 & 19.75 & \textbf{1722.2*} \\
    7 & 13.25 & 19.19 & \textbf{1862.4*} & 7 & 19.75 & 22.91 & 1135.9 \\
    7 & 19.53 & 22.42 & 1028.8 & & & & \\
    8 & 1.806 & 4.386 & 810.6  & 8 & 1.200 & 4.800 & 819.5 \\
    8 & 4.386 & 10.96 & \textbf{2453.3*} & 8 & 4.800 & 10.20 & \textbf{2550.0*} \\
    8 & 10.96 & 15.35 & 1775.4 & 8 & 10.20 & 15.60 & \textbf{1571.2*} \\
    8 & 16.51 & 20.51 & 1333.7 & 8 & 16.80 & 22.80 & 1782.1 \\
    8 & 20.51 & 22.70 & \textbf{922.7*}  & 8 & 22.80 & 24.60 & 708.2 \\
    8 & 24.76 & 27.47 & 676.9  & & & & \\
    9 & 0.000 & 4.164 & \textbf{1509.6*} & & & & \\
    9 & 4.164 & 7.192 & \textbf{1180.6*} & 9 & 0.000 & 7.630 & 3241.7 \\
    9 & 7.823 & 11.60 & \textbf{1392.5*} & 9 & 7.630 & 10.56 & 1083.7 \\
    9 & 11.60 & 15.52 & \textbf{1145.8*} & 9 & 11.73 & 13.49 & 736.6 \\
    9 & 17.41 & 19.18 & \textbf{646.0*}  & 9 & 15.26 & 17.02 & 536.6 \\
    9 & 19.18 & 21.07 & 750.7  & 9 & 17.02 & 21.71 & 1374.2 \\
    10 & 0.369 & 3.447 & \textbf{1071.2*} & 10 & 0.000 & 3.436 & 1306.2 \\
    10 & 3.447 & 6.156 & 975.4  & 10 & 3.436 & 6.299 & 1143.2 \\
    10 & 6.156 & 8.618 & 984.9  & 10 & 6.299 & 9.735 & \textbf{969.1*} \\
    10 & 8.618 & 10.09 & 556.3  & & & & \\
    11 & 0.258 & 4.515 & 1622.4 & 11 & 0.000 & 4.800 & 1939.5 \\
    11 & 4.515 & 9.417 & \textbf{1718.9*} & 11 & 4.800 & 10.20 & 1504.7 \\
    11 & 18.96 & 24.63 & 2069.9 & 11 & 18.60 & 27.00 & 3116.1 \\
    12 & 0.121 & 1.274 & 387.2 & & & & \\
    12 & 1.274 & 3.096 & \textbf{830.9*}  & 12 & 0.000 & 1.129 & 532.8 \\
    12 & 3.096 & 4.128 & 346.4 & & & & \\
    12 & 4.128 & 5.524 & \textbf{571.9} & 12 & 1.129 & 5.364 & \textbf{1485.1*} \\
    12 & 5.524 & 6.799 & 451.6 & & & & \\
    12 & 6.799 & 8.316 & 534.0  & 12 & 5.364 & 9.317 & \textbf{1633.6*} \\
    12 & 8.316 & 9.834 & \textbf{738.7*} & & & & \\
    13 & 0.081 & 2.042 & \textbf{560.5*} & 13 & 0.000 & 1.159 & 589.8 \\
    13 & 2.042 & 3.185 & 382.2 & & & &  \\
    13 & 3.349 & 4.084 & 263.3 & 13 & 4.179 & 6.838 & 977.6 \\
    13 & 4.084 & 6.371 & \textbf{795.4*} & & & & \\
    13 & 7.270 & 9.149 & 582.1 & 13 & 7.598 & 9.498 & 466.6 \\
    13 & 9.149 & 10.04 & 400.0 & & & & \\
    13 & 10.04 & 11.51 & 574.9 & 13 & 9.498 & 11.77 & 967.4 \\
    14 & 0.000 & 1.733 & \textbf{685.2*} & 14 & 0.000 & 2.121 & 571.6 \\
    14 & 1.915 & 3.739 & 478.4 & & & & \\
    \bottomrule
  \end{tabular}
\end{table}

\end{document}